\documentstyle[12pt,epsf]{article}

\newcommand{\be}{\begin{equation}}
\newcommand{\ee}{\end{equation}}

\begin{document}

\bibliographystyle{plain}

\title{\vspace*{-3cm}Hadron Spectrum in QCD  with Valence Wilson Fermions
  and Dynamical Staggered Fermions at $6/g^2=5.6$}

\author{
  \\
  {\small Khalil M. Bitar, R.G.~Edwards, U.M.~Heller, A.D.~Kennedy}\\[-0.2cm]
  {\small\it SCRI, The Florida State University, Tallahassee, FL 32306-4052,
             USA} \\[-0.2cm]
  \\[-0.2cm]  \and
  {\small Thomas~A.~DeGrand}\\[-0.2cm]
  {\small\it Physics Department, University of Colorado, Boulder, CO 80309,
             USA} \\[-0.2cm]
  \\[-0.2cm]  \and
  {\small Steven~Gottlieb, A.~Krasnitz} \\[-0.2cm]
  {\small\it Department of Physics, Indiana University, Bloomington, IN 47405,
             USA} \\[-0.2cm]
  \\[-0.2cm]  \and
  {\small J.~B.~Kogut} \\[-0.2cm]
  {\small\it Department of Physics} \\[-0.2cm]
  {\small\it University of Illinois, 1110 W. Green St., Urbana, IL 61801,
             USA} \\[-0.2cm]
  \\[-0.2cm]  \and
  {\small W.~Liu, Pietro~Rossi} \\[-0.2cm]
  {\small\it Thinking Machines Corporation, Cambridge, Mass. 02142,
             USA}\\[-0.2cm]
  \\[-0.2cm]  \and
  {\small Michael~C.~Ogilvie} \\[-0.2cm]
  {\small\it Department of Physics, Washington University,
             St. Louis, MO 63130, USA}\\[-0.2cm]
  \\[-0.2cm]  \and
  {\small R.~L.~Renken} \\[-0.2cm]
  {\small\it Department of Physics, University of Central Florida,
             Orlando, FL 32816, USA} \\[-0.2cm]
  \\[-0.2cm]  \and
  {\small D.~K.~Sinclair} \\[-0.2cm]
  {\small\it HEP Division} \\[-0.2cm]
  {\small\it Argonne National Laboratory, 9700 S. Cass Ave.,
             Argonne, IL 60439, USA}  \\[-0.2cm]
  \\[-0.2cm]  \and
  {\small R.~L.~Sugar} \\[-0.2cm]
  {\small\it Department of Physics, University of California, Santa Barbara,
             CA 93106, USA}  \\[-0.2cm]
  \\[-0.2cm]  \and
  {\small D.~Toussaint} \\[-0.2cm]
  {\small\it Department of Physics, University of Arizona, Tucson, AZ 85721,
             USA} \\[-0.2cm]
  \\[-0.2cm]  \and
  {\small K.~C.~Wang} \\[-0.2cm]
  {\small\it Department of Physics, University of New South Wales,
             Kensington, NSW 2203, Australia}  \\[-0.2cm]
  \\
}

\date{}
\maketitle
\thispagestyle{empty}   
\clearpage

\rightline{FSU-SCRI-92-61}
\rightline{COLO-HEP-278}
\rightline{hep-lat/9204008}

\vspace{2.0in}

\begin{abstract}
We present an analysis of hadronic spectroscopy for Wilson valence quarks
with dynamical staggered fermions at lattice coupling $6/g^2 = \beta=5.6$
at sea quark mass $am_q=0.01$ and 0.025,
and of Wilson valence quarks in quenched approximation at $\beta=5.85$
and 5.95, both  on $16^3 \times 32$ lattices.
We make comparisons with our previous results with dynamical staggered
fermions at the same parameter
values but on $16^4$ lattices doubled in the temporal direction.
\end{abstract}

\clearpage

\section{Introduction}

Calculations of hadron spectroscopy remain an important part of
nonperturbative studies of QCD using lattice methods.
(For a review of recent progress in this
field, see Ref. \cite{SPECTRUMREVIEWS}.)
We have been engaged in an extended program of calculation of the masses
and other parameters of the light hadrons in simulations which include
the effects of two flavors of light dynamical quarks.  These quarks
are realized on the lattice as staggered fermions. We have carried
out simulations with lattice valence quarks in both the staggered
and Wilson formulations.  Our reasons for performing simulations with
Wilson valence quarks is twofold: First, we are interested in seeing if
there are any effects of sea quarks on the spectroscopy of systems
containing either realization of valence quark. Thus this work complements
our parallel studies of spectroscopy with staggered valence and sea quarks,
and of spectroscopy with Wilson valence and sea  quarks.  Second, we are
interested in exploring the effects of sea quarks on simple matrix elements
such as the pseudoscalar meson decay constant. Most previous work has been
done with Wilson valence quarks in quenched approximation.
We consider that mixing the two realizations is not inappropriate
for a first round  of numerical simulations.

These simulations are performed on $16^3 \times 32$ lattices at lattice
coupling $6/g^2 = \beta=5.6$ with two masses of dynamical staggered
fermions, $am_q=0.025$ and $am_q=0.01$.  These are the same parameter
values as we used in our first round of simulations\cite{HEMCGC}
However, the first set of simulations have two known inadequacies. The
first is that most of our runs were carried out on lattices of spatial
size $12^3$. A short run on $16^3$ lattices with dynamical quark mass
$0.01$ showed that these lattices were too small:  baryon masses fell
by about fifteen per cent on the larger lattice compared to the smaller
one. Thus the $am_q=0.025$ results from Ref. \cite{HEMCGC} are suspect
and need to be redone.  We also felt that we needed more statistics on
the $am_q=0.01$ system.

Second, nearly all of our running was done on lattices of size $12^4$
or $16^4$; these lattices were doubled in the temporal direction to
$12^3 \times 24$ or $16^3\times 32$ for spectroscopy studies.  Doubling
the lattice introduced strange structure in the propagators of some of
the particles: the pion mass, in particular, showed strange oscillatory
behavior as a function of position on the lattice.  This behavior is
almost certainly due to doubling the lattice\cite{ALEX} and the best
way to avoid this problem is to begin with a larger lattice in the
temporal direction.

Finally, it is an open question how much sea quarks affect the hadronic
spectrum. In order to address this question, we have performed a set
of simulations in  the quenched approximation at lattice couplings
$\beta=5.85$ and 5.95, also on $16^3 \times 32$ lattices. As the reader
will see, our quenched results are rather similar to our results with
dynamic fermions; apparently at the particular values of sea quark mass
and lattice coupling where our simulations were performed, the effects
of sea quarks can be absorbed into renormalizations of the lattice
coupling and valence quark mass.

Some of the results described here have been presented in preliminary
form in Ref.~\cite{LAT9X}.  Several other papers which we are preparing
for publication complement the results presented here: we are also
preparing a paper on spectroscopy results with valence staggered
quarks, an analysis of simple matrix elements with Wilson valence
quarks, a study of valence quark Coulomb gauge wave functions, and a
study of glueballs and topology in the presence of dynamical staggered
quarks.

\section{The simulations}

Our simulations were performed on the Connection Machine CM-2 located
at the Supercomputer Computations Research Institute at Florida State
University.

We carried out simulations with two flavors of dynamical staggered
quarks using the Hybrid Molecular Dynamics algorithm\cite{HMD}
The lattice size is $16^3 \times 32$ sites and the lattice coupling
$\beta=5.6$. The dynamical quark mass is $am_q=0.01$ and
0.025. The total simulation length was 2000 simulation time units
(with the normalization of Ref. \cite{HEMCGC}) at each quark mass value,
after thermalization. The $am_q = 0.01$ run started from an equilibrated
$16^4$ lattice of our previous runs on the ETA-10, that was doubled in the
time direction and then re-equilibrated for 150 trajectories. The $am_q =
0.025$ run was started from the last configuration of the smaller mass run,
and then thermalized for 300 trajectories.
We recorded lattices
for the reconstruction of spectroscopy every 20 HMD time units, for a total
of 100 lattices at each mass value.

We computed spectroscopy with staggered sea quarks at five values of the
Wilson quark hopping parameter: $\kappa=0.1600$, 0.1585, 0.1565, 0.1525,
0.1410, and 0.1320. The first three values are rather light quarks
(the pseudoscalar mass in lattice units
ranges from about 0.25 to 0.45) and the other
three values correspond to heavy quarks (pseudoscalar masses of
0.65 to 1.5). The first three values are the ones we used in our first
round of experiments.
We computed masses of mesons with all possible combinations of
quark and antiquark mass; we only computed the masses of baryons made
of degenerate mass quarks.

The quenched simulations were performed at  lattice couplings of
$\beta=5.85$ and $5.95$, also on $16^3 \times 32$ lattices. They used
the Hybrid Monte Carlo algorithm\cite{HMC}
These simulations had a total length of 3200
time units at $\beta=5.85$ and 3800 time units at $\beta=5.95$. We
recorded lattices for Wilson valence spectroscopy every 40 time units,
for a sample of about 90 lattices at each coupling.
Our quenched spectroscopy was done at hopping parameter values designed to
reproduce as accurately as possible our earlier $\beta=5.6$ running:
we used $\kappa=0.1585$ and 0.1600 at $\beta=5.85$ and $\kappa=0.1554$
and 0.1567 at $\beta=5.95$. We only computed spectroscopy for
hadrons with degenerate quarks.

For the spectroscopy we used periodic
boundary conditions in all four directions of the lattice.  Our
lattices are long enough in the time direction and our interpolating fields
are good enough that we are always able to extract an asymptotic
mass.  The use of open boundary conditions introduces edge effects which
are hard to quantify and we have chosen to avoid them in the current round
of simulations.

We calculated hadron propagators in the following way:  We fix gauge in
each configuration in the data set to lattice Coulomb gauge  using an
overrelaxation algorithm\cite{OVERRELAX} and use sources  for the
quarks which spread out in space uniformly over the simulation volume
and restricted to a single time slice ( ``wall''
sources\cite{WALLSOURCES}).
This source is nonzero only on sites which
form one checkerboard of the lattice (the sum of $x$ plus $y$ plus $z$
coordinates is an even number). Our inversion technique  is  conjugate
gradient with preconditioning via ILU decomposition by
checkerboards\cite{DEGRANDILU}. We used a fast matrix inverter written
in  CMIS (Connection Machine Instruction Set)\cite{LIU}.

We use both a spread out sink as well as a
``pointlike'' sink where all the quarks lines end on the same site.
The ``wall'' sink is identical to the source, but sums over all sites.
We label hadron
propagators with wall sources and point sinks as ``WP'' and those
with a wall source and a wall sink as ``WW''.
We  combine the
quark propagators into hadron propagators in an entirely conventional
manner.
For Wilson hadrons we use relativistic wave functions\cite{WILSONWAVES}.
The baryon wave functions are:
\begin{eqnarray}
\noalign{\hbox{Proton:}}
 \left|P,s\right>&=& (u C\gamma_5 d)u_s \nonumber\\
                 &=&(u_1 d_2 - u_2 d_1 + u_3 d_4 -u_4 d_3) u_s\nonumber\\
\noalign{\hbox{Delta:\ }}
\left|\Delta_1,s\right>&=& (u_1 d_2 + u_2 d_1 + u_3 d_4 + u_4 d_3) u_s
     \nonumber\\
\left|\Delta_2,s\right>&=& (u_1 d_3 - u_2 d_4 + u_3 d_1 - u_4 d_2) u_s
\label{eqn1}
\end{eqnarray}
We have measured meson correlation functions using spin structures
$\bar \psi \gamma_5 \psi$ and $\bar \psi \gamma_0\gamma_5 \psi$ for the pion
and
$\bar \psi \gamma_3 \psi$ and $\bar \psi \gamma_0\gamma_3 \psi$ for the rho,
which we refer to as ``kind = 1'' and ``kind = 2'' for the pseudoscalar
and vector, respectively.

To extract masses from the hadron propagators, we must average the
propagators over the ensemble of gauge configurations, estimate the
covariance matrix and use a fitting routine to get an estimate of the
model parameters.  The lattices used for Wilson spectroscopy with
staggered sea quarks are separated by 20 HMD time units and do not show
any discernable time correlations with each other.  The quenched
simulations, spaced 40 Hybrid Monte Carlo time units apart, show some
residual  time correlation when we compare the error on the pion
effective mass blocking various numbers of successive lattices
together.  We attempt to take these correlations into effect by
blocking three successive lattices together before fitting the data.

We use the full covariance matrix in fitting the propagators in order
to get a meaningful estimate of the goodness of fit.
Reference\cite{DOUGFIT} discusses this fitting procedure in detail.

\section{Spectroscopy Results}

We determined hadron masses by fitting our data under the assumption
that there was a single particle in each channel.  This corresponds to
fitting for one decaying exponential and its periodic partner. We
calculated effective masses by fitting two successive distances, and
also made fits to the propagators over larger distance ranges.

In selecting the distance range to be used in the fitting, we have
tried to be systematic.  We somewhat arbitrarily choose the best
fitting range as the range which maximizes the confidence level of the
fit (to emphasize good fits) times the number of degrees of freedom (to
emphasize fits over big distance ranges) divided by the statistical
error on the mass (to emphasize fits with small errors). We typically
restrict this selection to fits beginning no more than 11 or 12
timeslices from the origin.

\subsection{Simulations with sea quarks}

\begin{figure}
\epsfxsize=\columnwidth
\epsffile{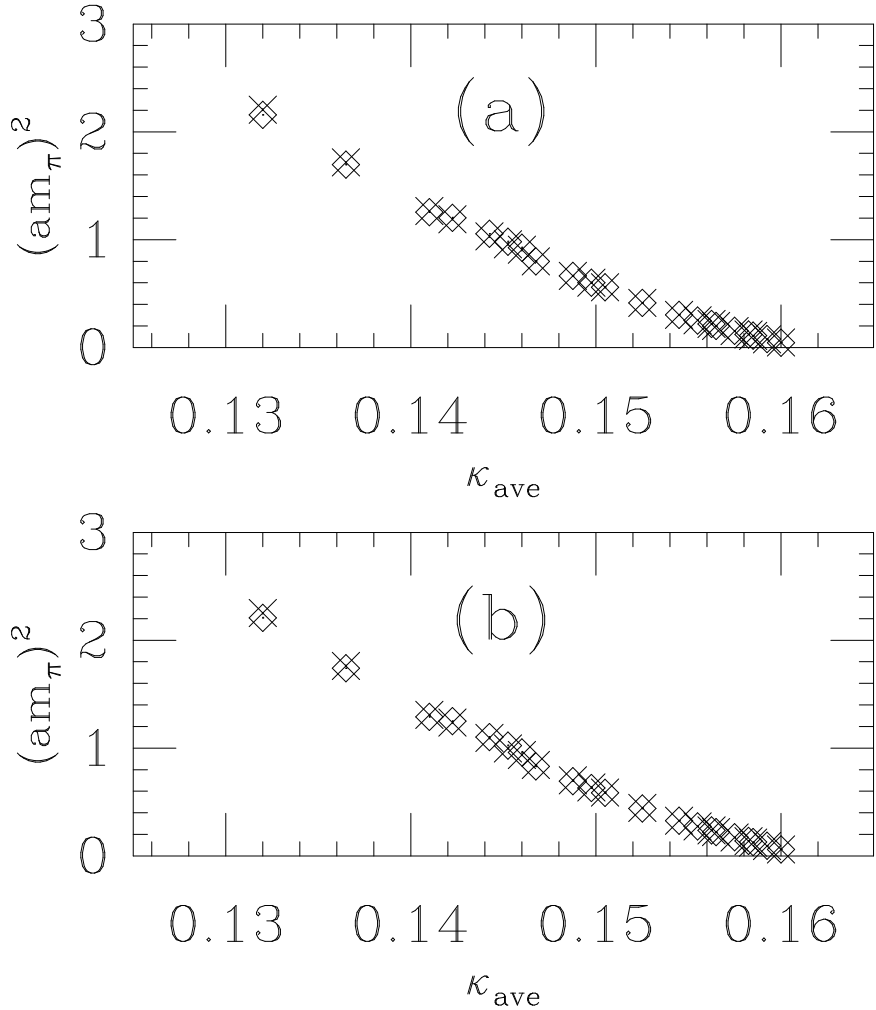}
\caption{
  Best fit masses (from fits to a range) for the pseudoscalar as a function of
  the average hopping parameter.
  Data are labelled by type (WP or WW)
  (described in the text)
  by crosses (WP) and  diamonds (WW).
  Figure (a) is for sea quark mass $am_q=0.01$, (b) for $am_q=0.025$.
}
\label{fig1}
\end{figure}

\begin{figure}
\epsfxsize=\columnwidth
\epsffile{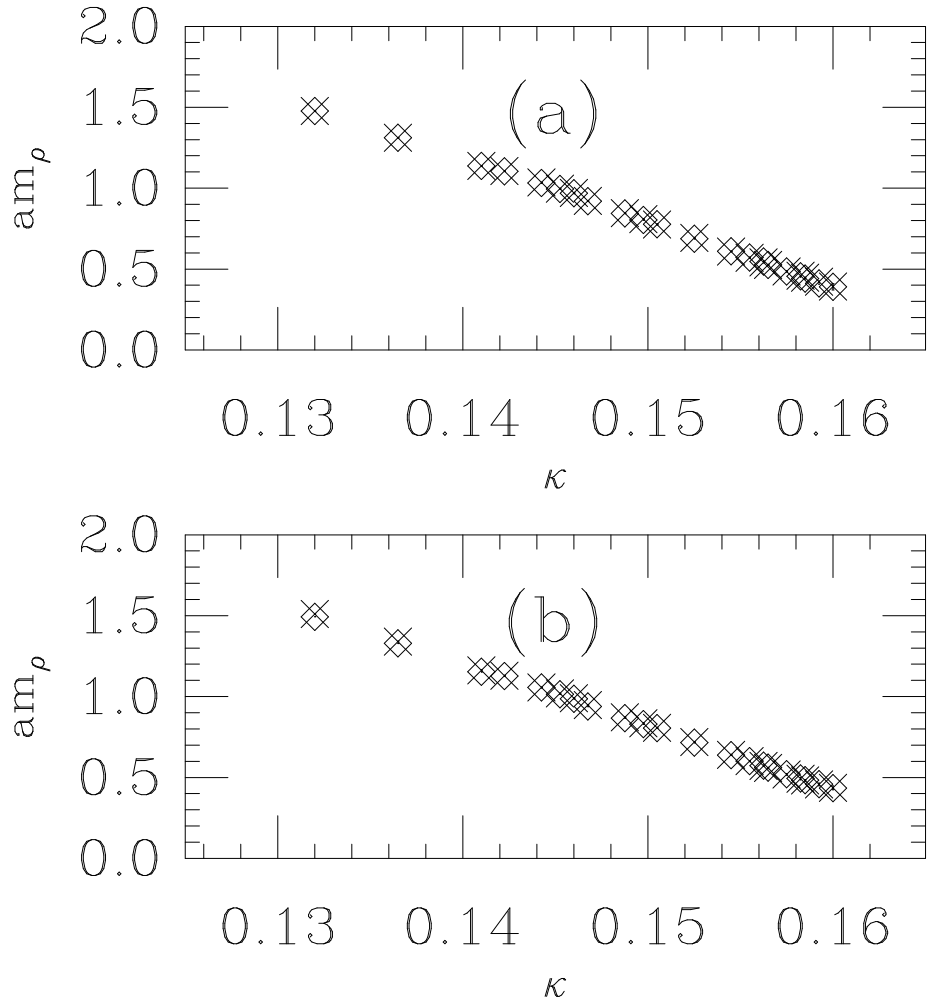}
\caption{
  Best fit masses (from fits to a range) for the vector meson as a function of
  the average hopping parameter.
  Figure (a) is for sea quark mass $am_q=0.01$, (b) for $am_q=0.025$.
}
\label{fig2}
\end{figure}

\begin{figure}
\epsfxsize=\columnwidth
\epsffile{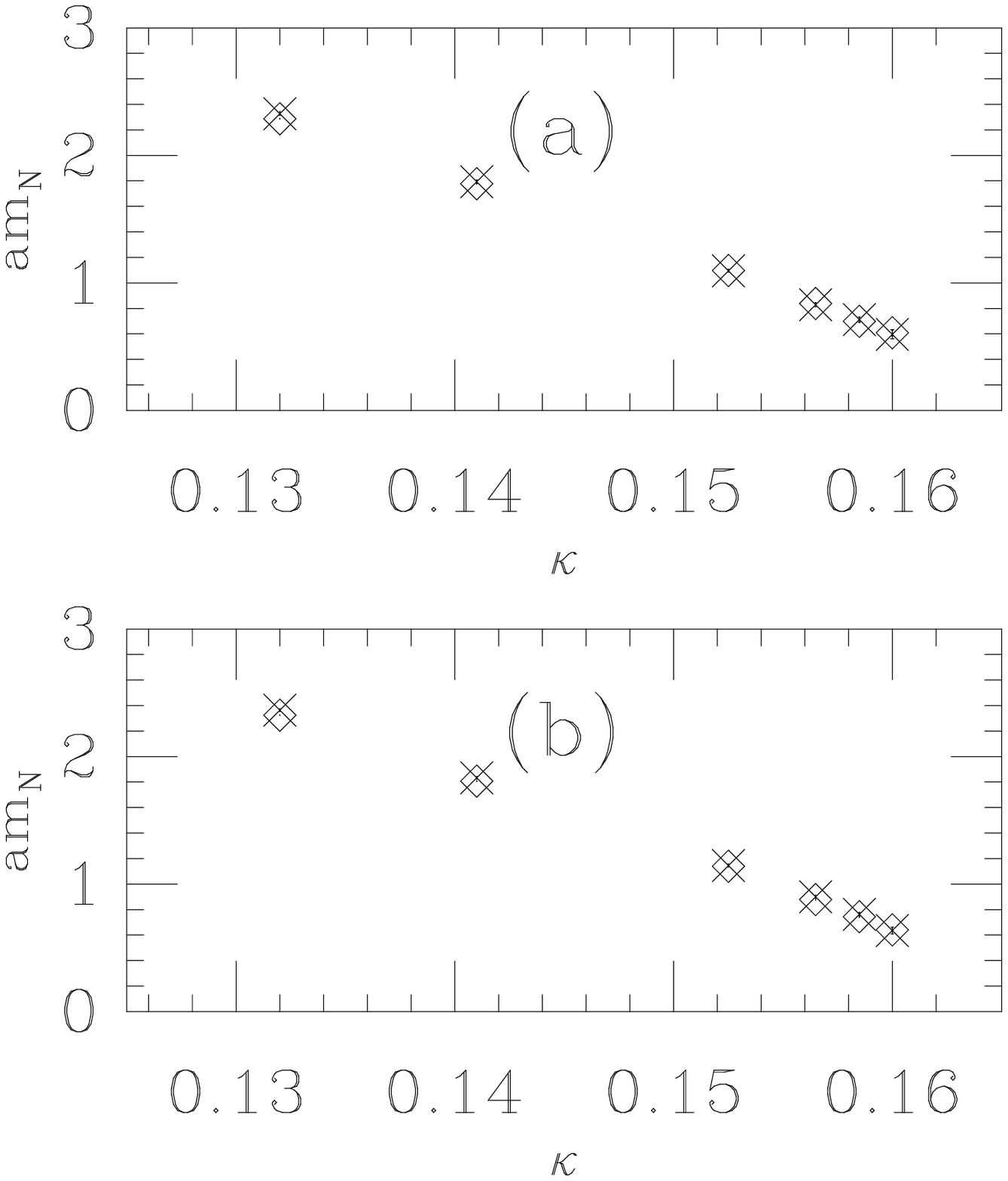}
\caption{
  Best fit masses (from fits to a range) for the proton as a function of
  hopping parameter.
  Figure (a) is for sea quark mass $am_q=0.01$, (b) for $am_q=0.025$.
}
\label{fig3}
\end{figure}

\begin{figure}
\epsfxsize=\columnwidth
\epsffile{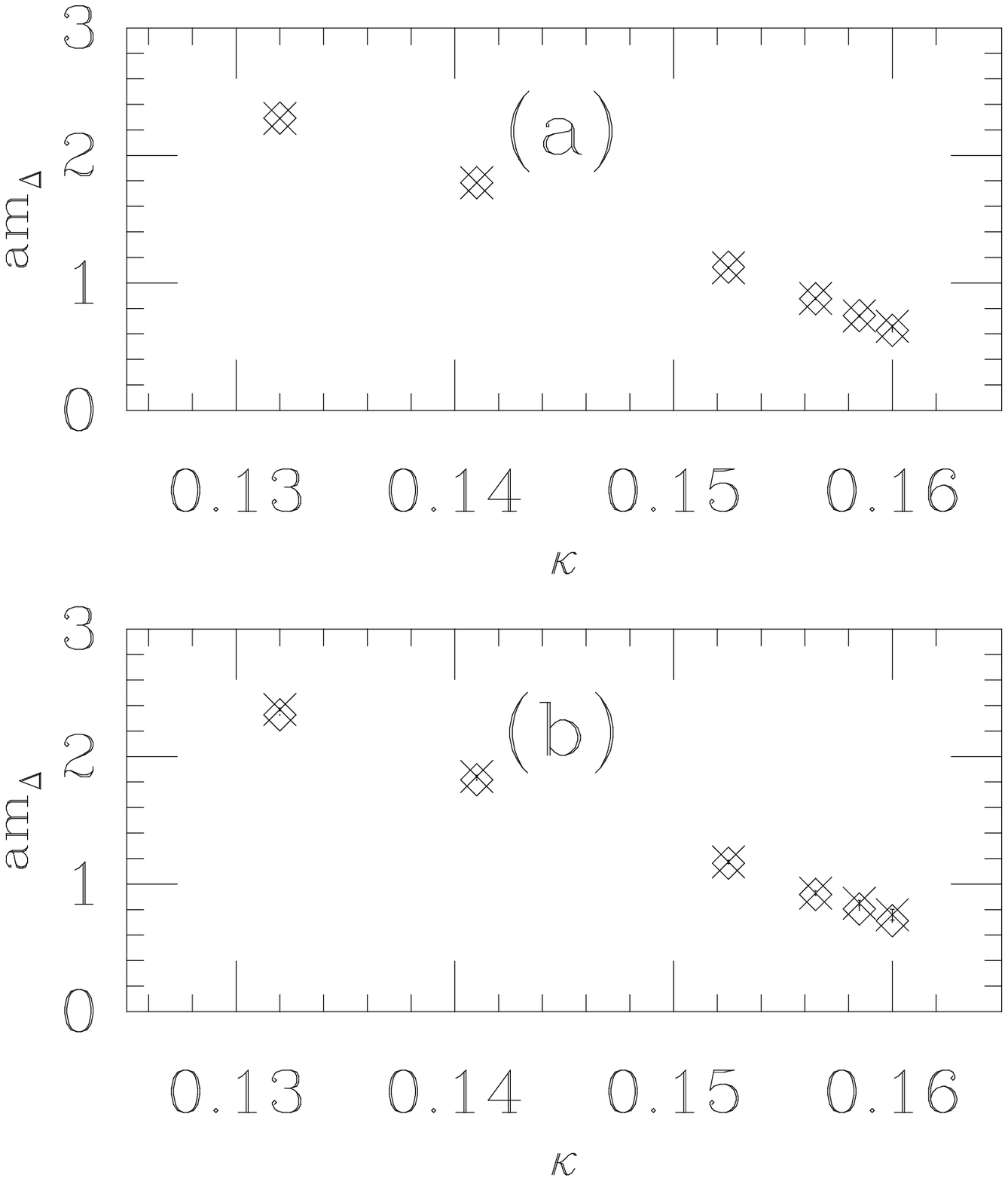}
\caption{
  Best fit masses (from fits to a range) for the delta as a function of
  hopping parameter.
  Figure (a) is for sea quark mass $am_q=0.01$, (b) for $am_q=0.025$.
}
\label{fig4}
\end{figure}

\begin{figure}
\epsfxsize=\columnwidth
\epsffile{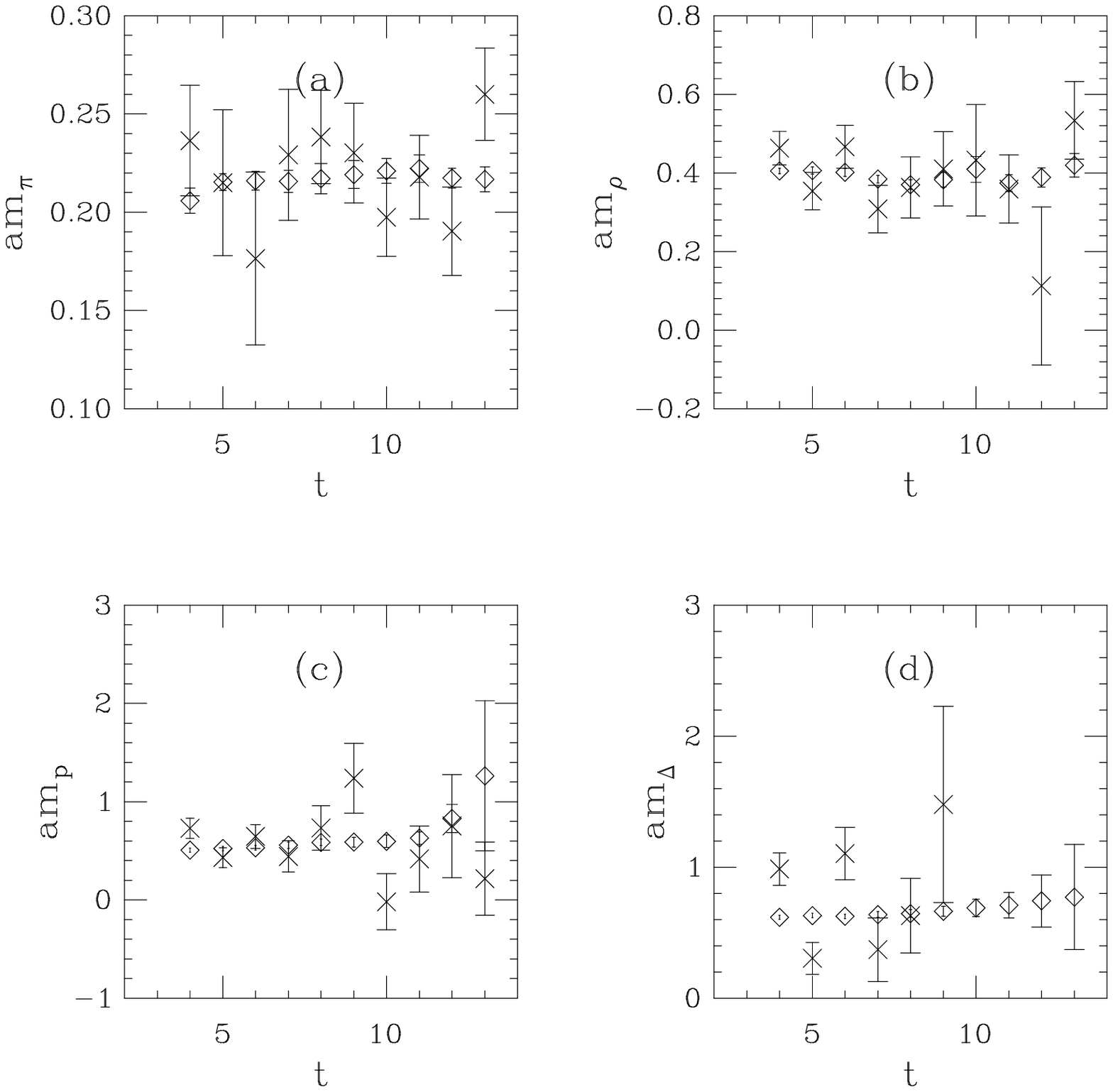}
\caption{
  Effective mass fits  to $\kappa=0.1600$  data: (a) pion, (b) rho,
  (c) proton, and (d) delta.  Data are labelled by type (WP or WW)
  by crosses (WP) and diamonds (WW).
}
\label{fig5}
\end{figure}

\begin{figure}
\epsfxsize=\columnwidth
\epsffile{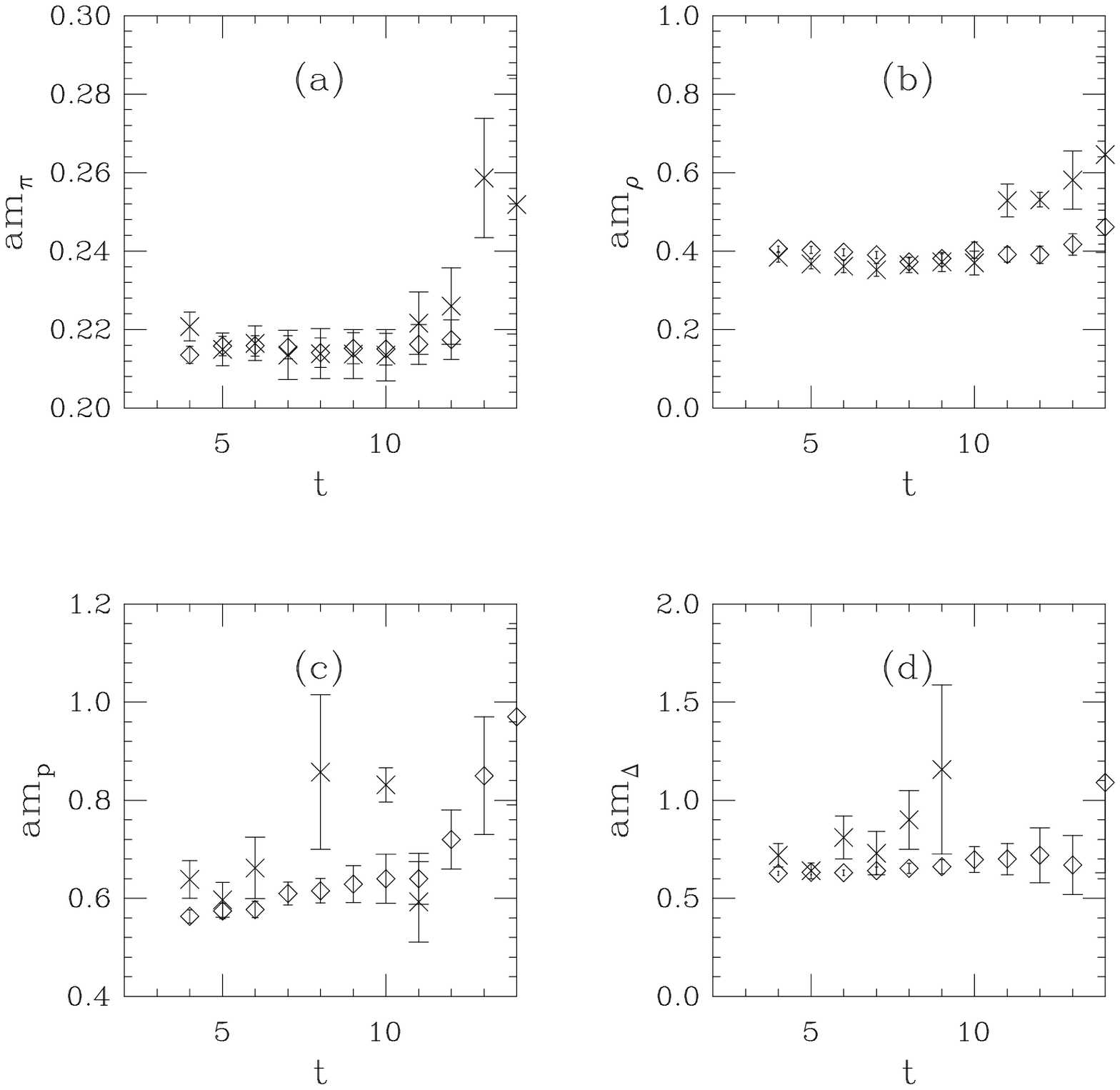}
\caption{
  Fits from $t=D_{min}$ to 16  to $\kappa=0.1600$ data: (a) pion, (b) rho,
  (c) proton, and (d) delta.  Correlator types are labelled as in
  Fig.~\protect\ref{fig1}-\protect\ref{fig5}.
}
\label{fig6}
\end{figure}

We computed spectroscopy for the six possible values of valence quark
hopping parameter given above, with both `WP' and `WW' correlation
functions.  We first show the global results to spectroscopy, giving
the best-fit value for the mass for each value of hopping parameters.
We display masses as a function of the average hopping parameter
${1 \over 2}(\kappa_1 + \kappa_2)$ for mesons and as a function of $\kappa$
for baryons (recall that for each sea quark mass we only studied
baryons in which all three quarks have the same mass) in
Figs.~\ref{fig1}--\ref{fig4}.  In all these figures masses are quoted in
lattice units.  We display  plots of effective mass in Fig.~\ref{fig5} and of
mass versus $D_{min}$ (with $D_{max}=16$) for $\kappa=.1600$  data in
Fig.~\ref{fig6}.  Since the "kind=1" and "kind=2" operators produce essentially
identical spectroscopy we only show "kind=1" results in these figures
in an attempt to avoid clutter. Best fit values for each particle are
shown in Tables~\ref{tab1}--\ref{tab8}.

For all but two cases the fitting was straightforward.
However, we had two difficult data sets,  $\kappa=.1320$ spectroscopy
and the $am_q=0.01$, $\kappa=.1600$ delta.

\begin{figure}
\epsfxsize=\columnwidth
\epsffile{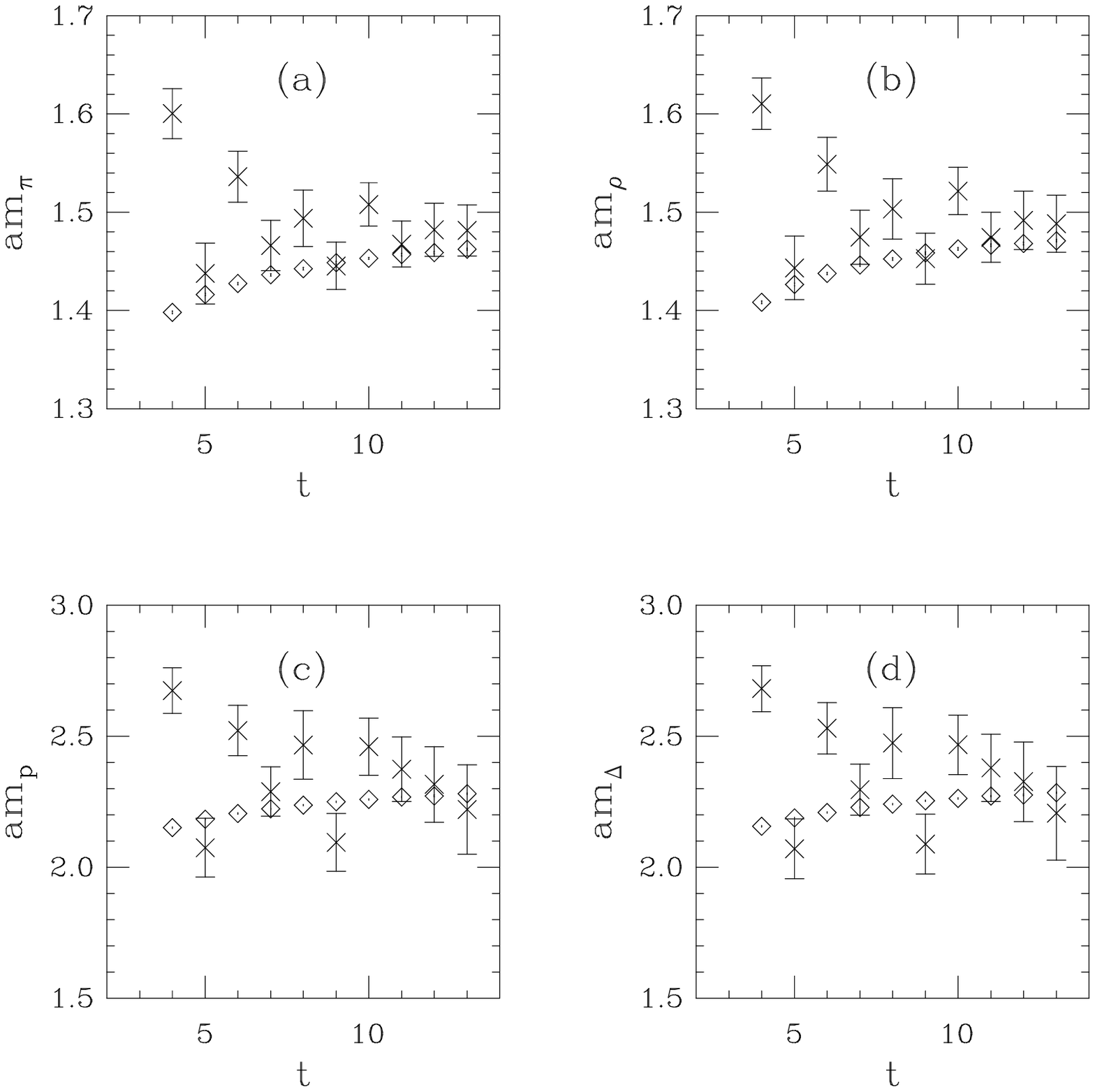}
\caption{
  Effective mass fits  to $\kappa=0.1320$  data: (a) pion, (b) rho,
  (c) proton, and (d) delta.  Data are labelled by type (WP or WW)
  by crosses (WP) and diamonds (WW).
}
\label{fig7}
\end{figure}

For all particles except those containing one or more of the heaviest
quarks both kinds of correlation functions give consistent results,
with the `WP' correlators typically showing the smallest uncertainty.
For mesons or baryons containing a heavy $\kappa=.1320$ quark, however,
the `WP' fits do not settle down to an asymptotic value.  The effective
mass drifts continuously with $t$ value and fits to a range have
unacceptably high chi-squared's. The `WW' fits are more acceptable
(have a chi-squared near one per degree of freedom).  Possibly what is
happening is this:  For these states the wall source has poor overlap
on the lightest hadron in a channel, since bound states of heavy quarks
have small spatial extent.  The WP correlators do not give a
variational bound and it happens that they approach an asymptotic mass
from below. The WW correlators  approach an asymptotic value from
above, but are noisier; one gets a statistically more acceptable fit
because of larger uncertainties on the individual points.  As an
example, we show effective masses as a function of $t$ for the
$ma=0.01$, $\kappa=0.1320$ data in Fig.~\ref{fig7}.

Next, the $am=0.01$, $\kappa=.1600$ delta mass is considerably lighter
than we saw in our old running and nearly degenerate with the nucleon.
At this light valence quark mass the `WW' operators are so noisy that
they are useless, but the same behavior is seen in  `WP'  delta
operators with both spin structures of Eqn.~\ref{eqn1}.
A comparison of the two
operators is shown in Fig.~\ref{fig5}d (effective masses)
and \ref{fig6}d (fits to a range $d_{min}$ to 16).
The signal for a light delta appears to be
stable for fit distances ranging from $d_{min}=4$ out to 8 or 9, and
then the signal deteriorates so rapidly that we cannot trust our fits.
We do not know of a reason for this effect. A coding error would mix
some component of the nucleon into the delta and the asymptotic mass in
the delta channel would be the nucleon's.  However, the $am_q=0.025$
delta is  measurably heavier than the nucleon, which argues against a
coding error.  One should note that the `WP' correlators are not
variational.  It is  possible that we are seeing a signal attempting to
approach an asymptotic value from below, which becomes lost in the
noise before its asymptotic value is reached.

Our data can be compared with our previous $16^4$ runs:  The
$\kappa=.1585$ and .1600 pseudoscalar vector, and proton masses are
consistent with the old numbers.  The $\kappa=.1585$ deltas are
consistent but the new $\kappa=.1600$ delta is quite a bit lighter
(0.63 vs. 0.74).   At $\kappa=.1565$ we had run before only on $12^3$
lattices. The new proton is lighter (0.84 vs. 0.89) and so is the delta
(0.87 vs. 0.96).  Clearly the old $\kappa=.1565$ masses were
compromised by the size of the simulation volume (as were the
simulations at lighter valence quark mass). All $am_q=0.025$ baryons
are also ten to fifteen per cent lighter than their values on $12^3$
lattices.

We do not see any oscillations in the pion effective mass at
$\kappa=.1600$ which we saw in the old doubled $16^4$ running
(compare Fig.~\ref{fig5}.)
There, however, the effect was most dramatic for the
staggered valence quark pion.

Assuming that $m_\pi^2$ is linear in $\kappa$ (as we expect from
current algebra considerations), we can compute the critical coupling
$\kappa_c$ at which the pion becomes massless.
We extrapolate using
\be
(m_\pi a)^2 = A({1 \over \kappa} - {1 \over \kappa_c})
\label{eqn2}
\ee
The fit is acceptable only for the three
lightest quark masses and the final numbers are essentially unchanged
whether we use all six combinations of quarks in the pseudoscalar or
restrict ourselves to the three cases of degenerate quark masses.
We find that
$A=1.10(1)$  and $\kappa_c= 0.1610(1)$ for $am_q=0.01$.  These numbers
are in good agreement with our previous results ($A=1.15(16)$ and
$\kappa=0.1611(1)$).

\begin{figure}
\epsfxsize=\columnwidth
\epsffile{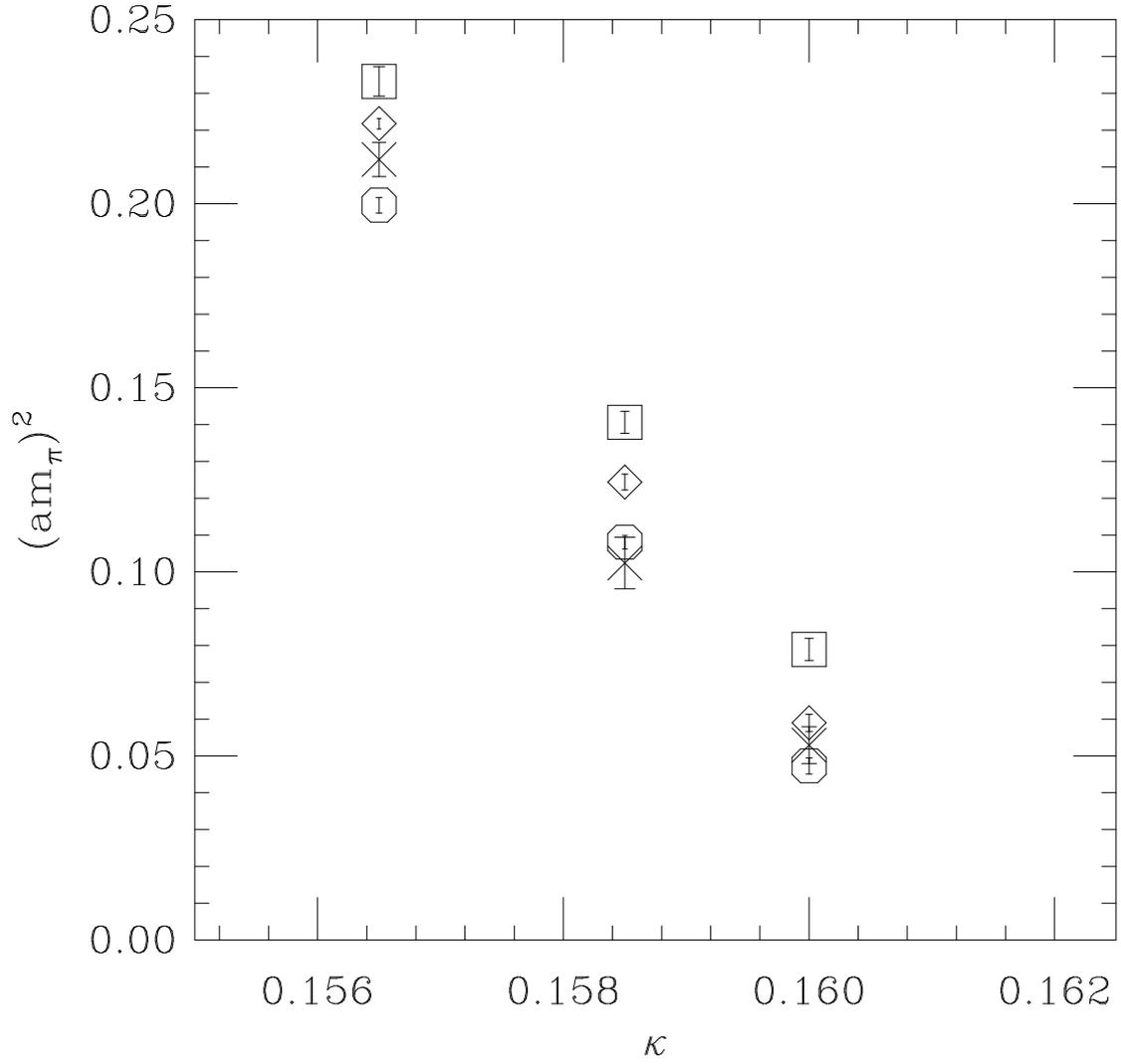}
\caption{
  Square of pion mass versus hopping parameter from
  the old $am_q=0.025$ data (diamonds), new $am_q=0.025$ data (squares),
  old $am_q=0.01$ data (crosses) and new $am_q=0.01$ data (octagons)
}
\label{fig8}
\end{figure}

The $am_q=0.025$ numbers are quite different from our previous study.
There we had $A= 1.15(16) $ and $\kappa_c=0.1618(1) $.
Here we have
$A=1.14(1)  $, $\kappa_c=0.1613(1)$ from the 'kind=1, WP' operator,
$A=1.17(2)$, $\kappa_c=0.1613(1)$ from the 'kind=2, WP' operator,
using only equal  quark mass pions.  This is such a large change that it
cannot be due to a statistical fluctuation. When we graph the square of
the pion mass from the old simulations (on a $12^3$ lattice) and
from the new simulations (on a $16^3$ lattice) we see that the new
pions are consistently lighter than the old ones.
The situation is illustrated in Fig.~\ref{fig8}, where we also show the
old and new $am_q=0.01$ data. It is very strange that a finite size effect
(if that is what we are seeing) would be stronger for heavier dynamical
fermion mass.

\begin{figure}
\epsfxsize=\columnwidth
\epsffile{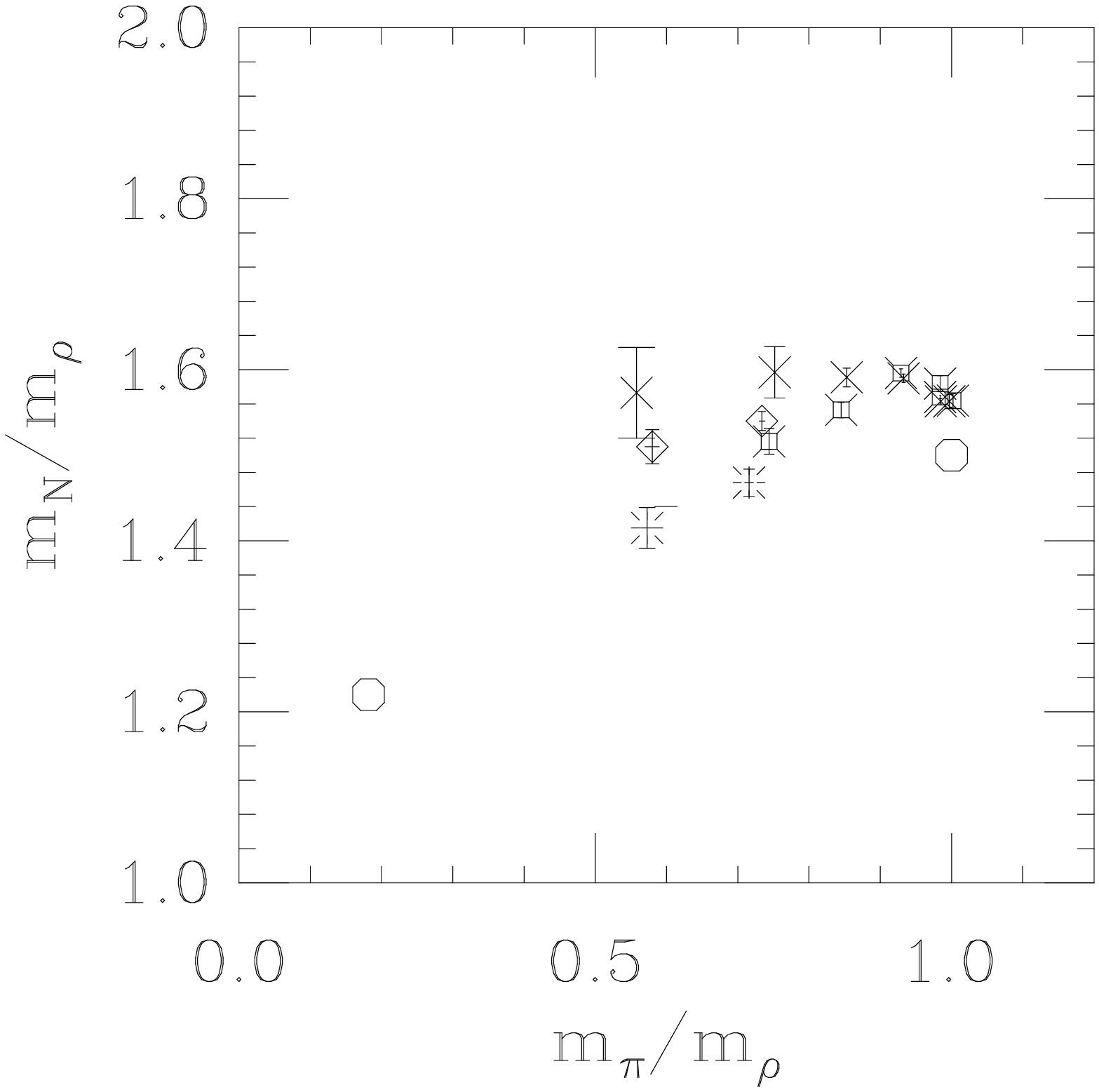}
\caption{
  Edinburgh plot for Wilson valence quarks.
  Data are:
  simulations with dynamical staggered fermions at $\beta=5.6$ and $am_q=0.01$
  from the $16^4$ running (diamonds) and from the $16^3\times 32$ running
  (crosses), $\beta=5.6$, $am_q=0.025$ $16^3 \times 32$ simulations
  (fancy squares),
  and quenched simulations at $\beta=5.85$
  and $\beta=5.95$ (bursts).  The circles show the expected infinite
  quark mass limit and the real-world point.
}
\label{fig9}
\end{figure}

\begin{figure}
\epsfxsize=\columnwidth
\epsffile{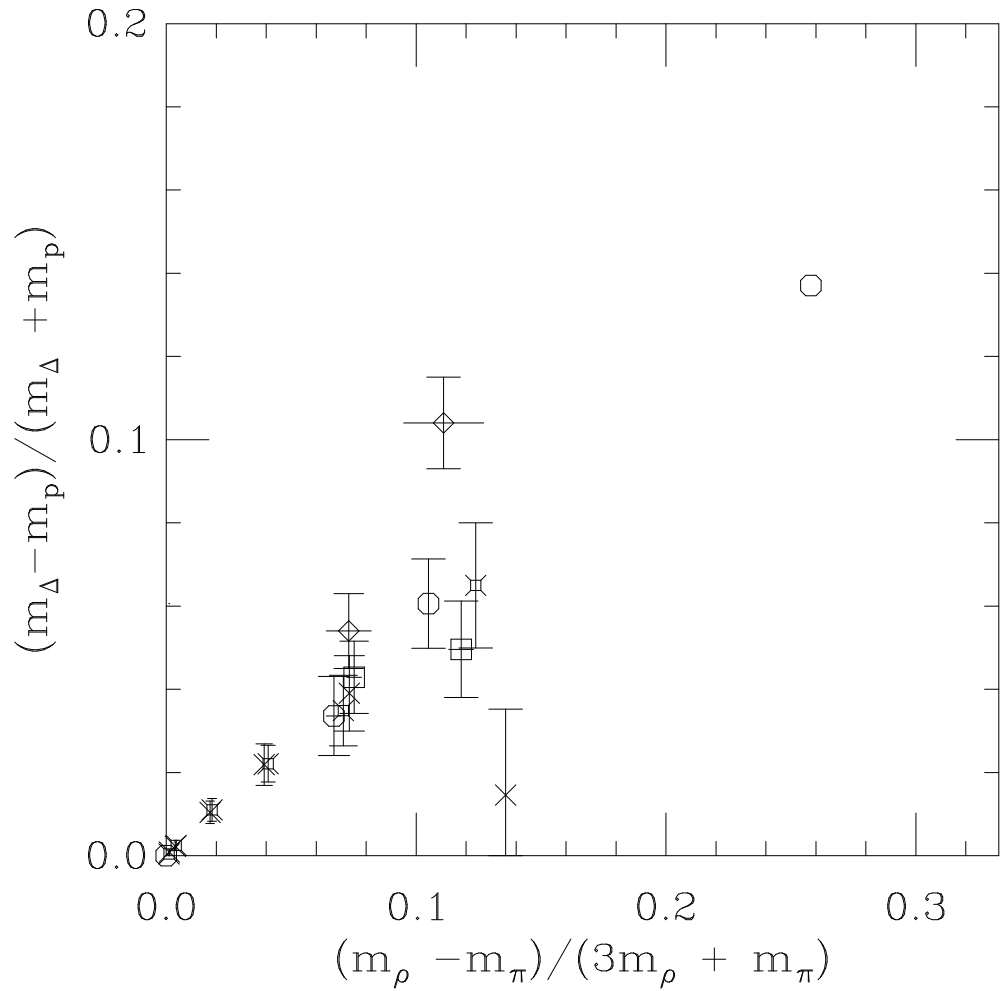}
\caption{
  Comparison of baryon and meson hyperfine splitting.  The two circles
  show the expected values of hyperfine splitting in the limit of infinite
  quark mass and from experiment.
  Points are labelled as in Fig.~\protect\ref{fig9}.
}
\label{fig10}
\end{figure}

In Fig.~\ref{fig9}
we present an Edinburgh plot ($m_p/m_\rho$ vs $m_\pi/m_\rho$).
This figure also includes data from other simulations we have performed.
Mass ratios, computed using correlated fits to a single exponential in
each channel, are shown in Tables~\ref{tab9}--\ref{tab12}.
We  quantify the magnitude
of hyperfine splittings in the meson and baryon sectors by comparing
the two dimensionless quantities
\be
R_M = {{m_\rho - m_\pi} \over {3 m_\rho + m_\pi}}
\label{eqn3}
\ee
and
\be
R_B = {{m_\Delta - m_p} \over {m_\Delta + m_p}}.
\label{eqn4}
\ee
Each of these quantities is the ratio of hyperfine splitting
in a multiplet divided by the center of mass of the multiplet.  A plot
of $R_M$ vs. $R_B$ is shown in Fig.~\ref{fig10}.

In the nonrelativistic quark model, the mass of a hadron is given by a sum
of constituent quark masses plus a color hyperfine term,
\be
M_H = \sum_i m_i + \xi_H \sum_{ij}{{\sigma_i \cdot \sigma_j}\over{m_im_j}}
\label{eqn5}
\ee
where $\xi$ is twice as great for mesons as for baryons because of
color\cite{DGG}
{}From this model one expects the ratio $R_B/R_M = 1$. For all but the lightest
quark mass data points, this is the behavior which our data shows.

If we wish to use our spectrum results to find a lattice spacing,
we can extrapolate particle masses to $\kappa_c$, fix one mass to
experiment, and use this mass to infer a lattice spacing.  We can do this
for the $\rho$, proton, and $\Delta$ for either sea quark mass.
Restricting our extrapolation to the lightest three valence hopping
parameters, we display our results in Table~\ref{tab13}.
Taking the rho as the particle whose mass is forced to its
physical value, we have an inverse lattice spacing of 2140 MeV or 2000 MeV,
proton masses of 1121   and  1116 MeV,
and deltas at 1198  and 1302 MeV.
Using the proton mass to set the scale, we have inverse lattice spacings
of 1800 or 1685 MeV,
rhos at 648  and 650  MeV
and deltas at 1008  and  1100  MeV.  In all cases the proton to rho
mass ratio is larger than experiment, and the proton-delta hyperfine
splitting is too small.


\subsection{Quenched Simulations}

In an attempt to
see whether any effects of dynamic fermions could be seen in the
spectroscopy, we performed a quenched simulation
with the same lattice volume as our  dynamical simulations
and with a large enough data set to overwhelm statistical
fluctuations\cite{LAT9X}

\begin{figure}
\epsfxsize=\columnwidth
\epsffile{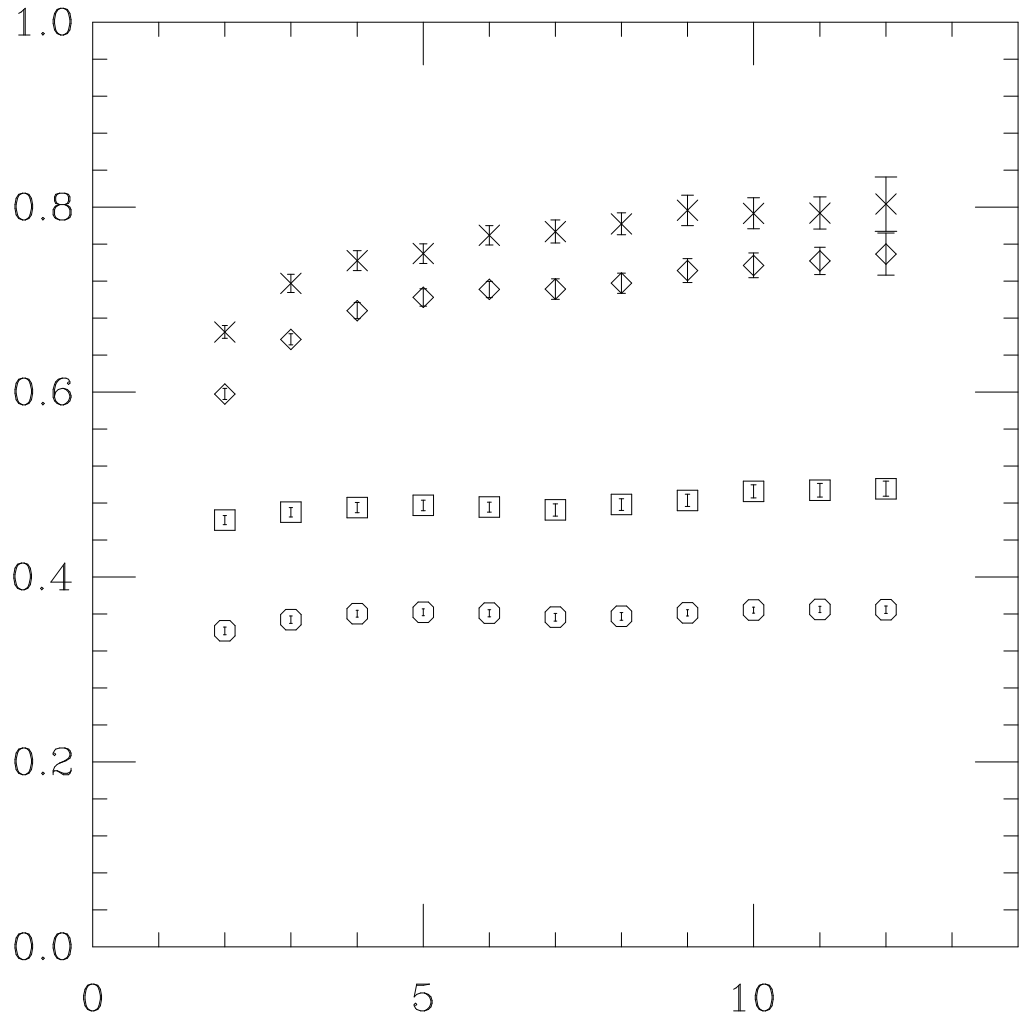}
\caption{
  Effective masses from WP operators from quenched spectroscopy
  at $\beta=5.95$, $\kappa=0.1554$. Particles in increasing order of mass
  are pion, rho, proton, and delta.
}
\label{fig11}
\end{figure}

\begin{figure}
\epsfxsize=\columnwidth
\epsffile{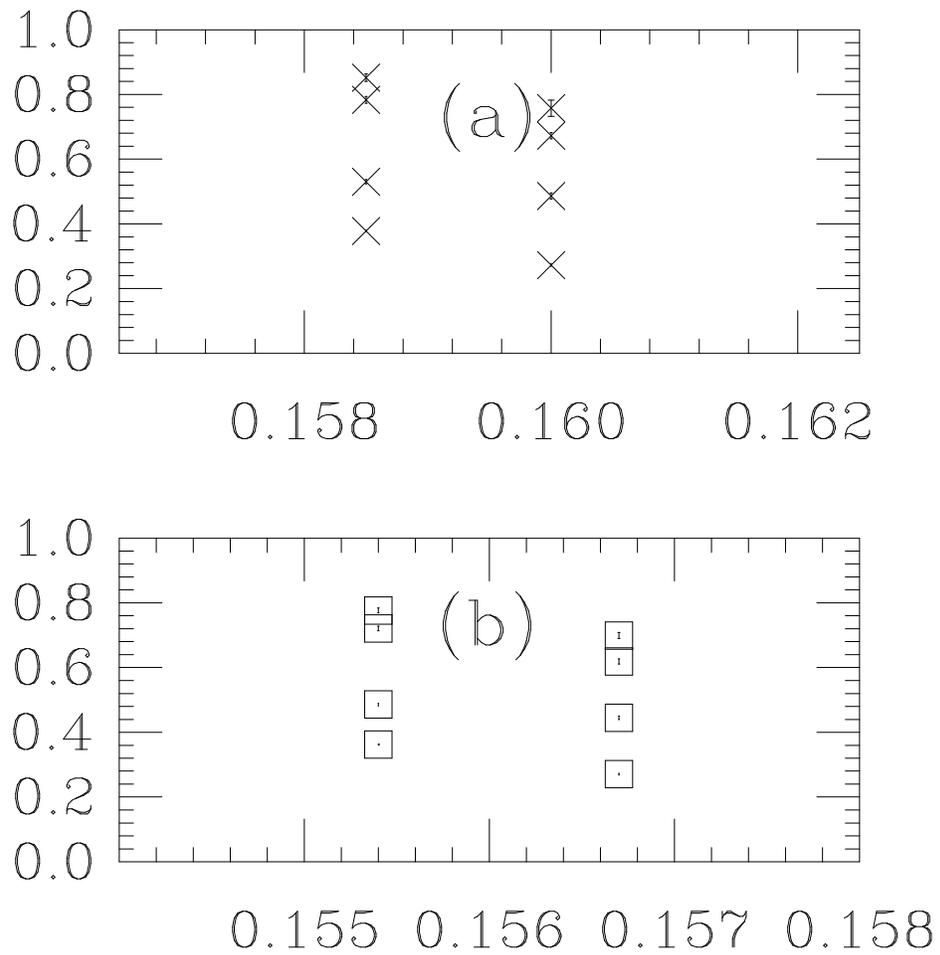}
\caption{
  Quenched masses from $\beta=5.85$ (a) and $\beta=5.95$ (b) simulations.
  Particles in increasing order of mass are pion, rho, proton, and delta.
}
\label{fig12}
\end{figure}

All fits are quite stable. We show one example of effective masses,
for the $\beta=5.95$, $\kappa=0.1554$ data set (Fig.~\ref{fig11}).
The best fits to a range of points, selected using the histogram
technique, begin at $t_{min}=6$ to 8, and are shown in Fig.~\ref{fig12} and
Tables~\ref{tab14} and \ref{tab15}.
Mass ratios are found in Tables~\ref{tab16} and \ref{tab17}.
Our quenched data at $\beta=5.85$, $\kappa=.1585$ are consistent within
statistical errors with the earlier work of Iwasaki, et. al.\cite{IWASAKI}

\begin{figure}
\epsfxsize=\columnwidth
\epsffile{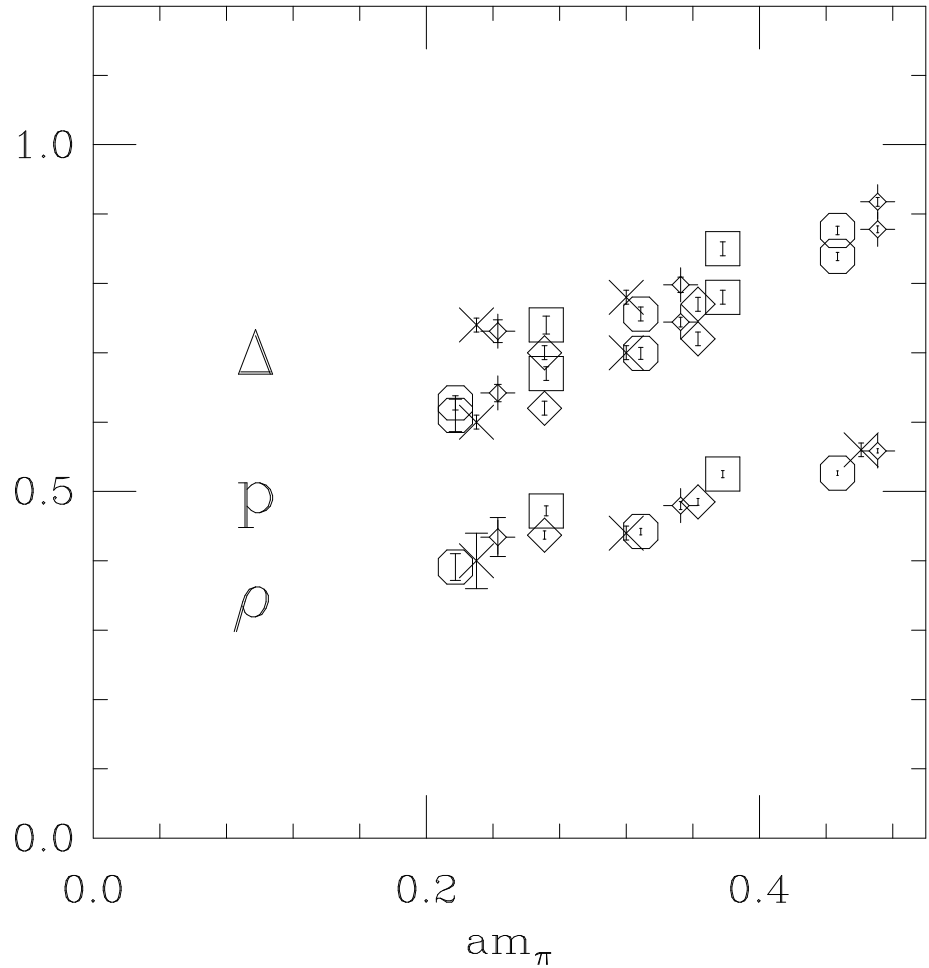}
\caption{
  Hadron masses  ($\rho$, p and $\Delta$) as a function of $m_\pi$
  for quenched and dynamical staggered simulations. Data are labelled
  with squares and diamonds for quenched $\beta=5.85$ and 5.95 simulations,
  crosses for the $16^4$ $am_q=0.01$ simulations, and octagons and fancy
  diamonds for the $am_q=0.01$ and 0.025 data presented in this paper.
}
\label{fig13}
\end{figure}

\begin{figure}
\epsfxsize=\columnwidth
\epsffile{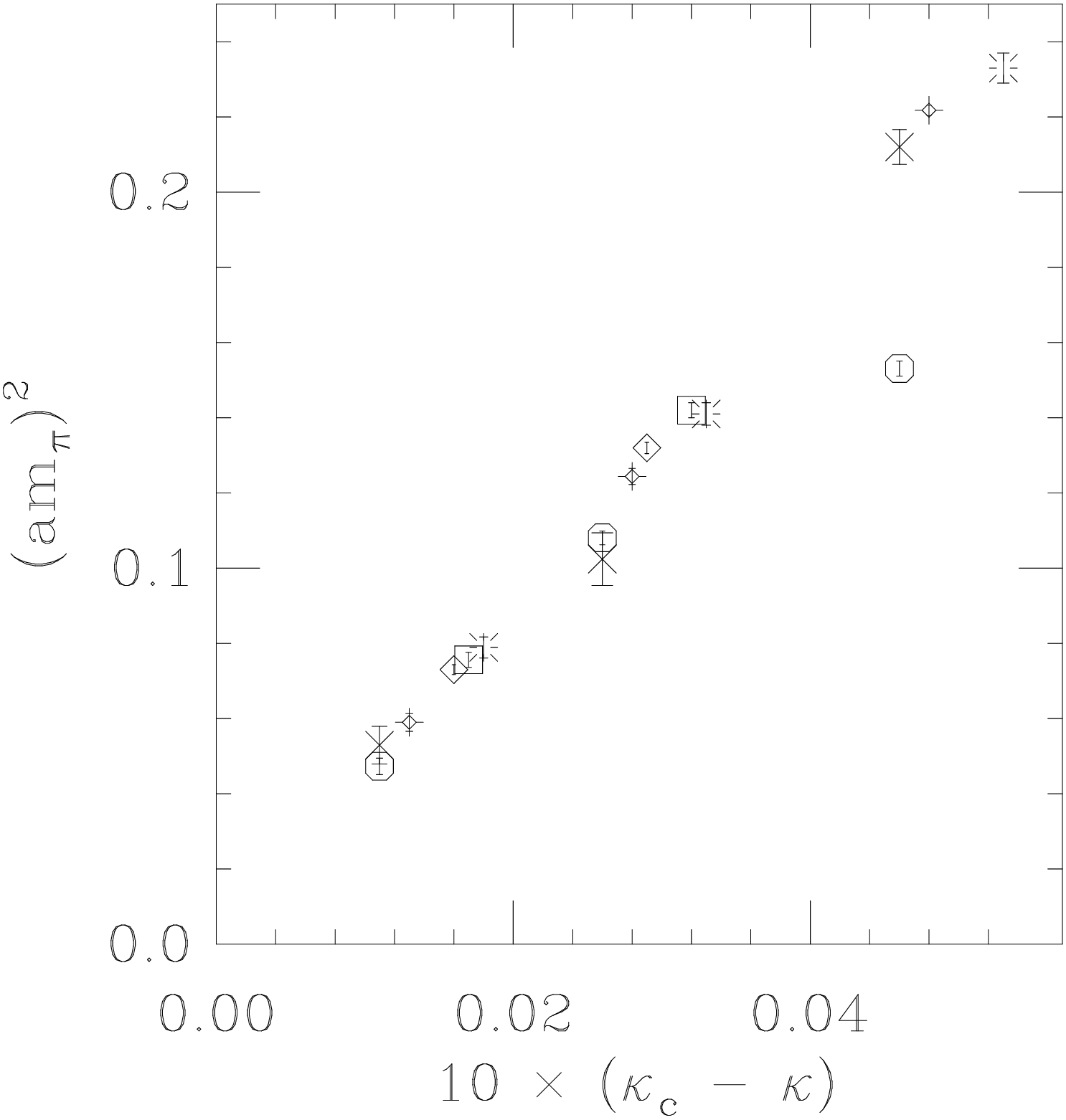}
\caption{
  Square of pion mass in lattice units
  for quenched and dynamical staggered simulations. Data are labelled
  as in Fig.~\protect\ref{fig13} and the $16^4$ $am_q=.025$ dynamical fermion
  data are labelled by a fancy diamond.
}
\label{fig14}
\end{figure}

While one cannot say anything about the behavior of the pion mass
as a function of hopping parameter with two data points per $\beta$ value,
one can still extrapolate the square of the pion mass to zero.  Doing so,
we find $\kappa_c = 0.1617(1)$, $A= 1.12(4)  $ for $\beta=5.85$ and
$\kappa_c = 0.1583(1)$, $A= 1.10(3)  $ for $\beta=5.95$.

Finally, as a direct way of displaying any differences between our dynamical
and quenched simulations, we show hadron masses (rho, proton, and delta)
as a function of the pion mass in lattice units (Fig.~\ref{fig13}).
With the possible
exception of the $am_q=0.01$, $\kappa=.1600$ delta, one cannot see any
strong difference between spectroscopy with or without dynamical fermions
at the parameter values used in this study. There is a hint in the
Edinburgh plot that the nucleon to rho mass ratio
in the presence of $am_q=0.01$
dynamical fermions is a bit higher
than in quenched approximation.
We display the square of
the pion mass in lattice units as a function of $\kappa_c - \kappa$
in Fig.~\ref{fig14}. Again all the data at the lightest valence quark masses
appear to lie on a (nearly) universal curve.

\section{Conclusions}

This concludes our program of  spectroscopy for Wilson valence quarks
with staggered sea quarks at $\beta=5.6$ on $16^3\times32$ lattices.
The spectroscopy we see is generally consistent with our earlier
results (when performed on lattices of the same spatial volume) and
represents an improvement over our previous results insofar as the
simulation volume is larger.  By comparing results from spatial volumes
of $12^3$ and $16^3$, we saw that for the smaller volume baryons with
light valence quarks suffer from finite lattice size effect regardless
of the dynamical fermion mass we used. This is another piece of
evidence which suggests that sea quark properties are much less
important than valence quark properties, in the parameter range we have
studied.  Perhaps we are also seeing the same effect on the pion mass.
Of course, one can not be sure that our results are still not
contaminated by finite volume effects on $16^3$ lattices; to test that
would require simulations in a larger volume with otherwise unchanged
parameter values.

When we compare our spectroscopy with dynamical fermions to quenched
results we do not see any dramatic differences. Apparently at the
parameter values of the simulation sea quarks simply do not affect
spectroscopy above the five to ten per cent level.

However, we have to say that we have always regarded this quark
combination as something done as much for expedience as for curiosity.
Wilson quarks and staggered quarks have very different symmetries.
The next barrier in spectroscopy calculations will occur when lattices
become large enough and quark masses become small enough that the
rho meson mass falls below the lightest $I=1$, $J=1$ $\pi\pi$ state
on the lattice, or that the lightest propagating state in the rho
channel is a $\pi\pi$ pair. These pions will each be made  of one of
the rho meson's valence quark(antiquark) and an antiquark(quark)
which has popped out of the
sea. It would be desirable if both quark and antiquark have the same internal
symmetry structure (for example, one will want to know the mass
of the pion in advance), and this argues against further use of
``hybrid'' quark calculations.

\section{Acknowledgments}

This work was supported by the U.~S. Department of Energy under contracts
DE--FG02--85ER--40213, 
DE--AC02--86ER--40253, 
DE--AC02--84ER--40125, 
W-31-109-ENG-38, 
and by the National Science Foundation under grants
NSF-PHY87-01775, 
NSF-PHY89-04035  
and
NSF-PHY86-14185. 
The computations were carried out at the
Florida State University Supercomputer Computations
Research Institute which is partially funded by the U.S. Department of
Energy through Contract No. DE-FC05-85ER250000.
We thank  T. Kitchens and J. Mandula for their
continuing support and encouragement.

\newpage

%
%
%

\begin{table}
\begin{center}
\begin{tabular}{|l|c|r|r|l|r|r|} \hline
\multicolumn{1}{|c|}{kind} & $\kappa_{ave}$ & $D_{min}$ & $D_{max}$
  & \multicolumn{1}{|c|}{mass} & $\chi^2$/dof & C.L.\\
\hline
 1 1 (WW)& 0.1320 & 5 & 16 & 1.485( 2) &     11.390/10  &      0.328 \\
 2 1 (WW)& 0.1365 & 5 & 16 & 1.310( 2) &      7.417/10  &      0.686 \\
 2 2 & 0.1410 & 11 & 16 & 1.120( 1) &      0.774/4  &      0.942 \\
 3 1 (WW)& 0.1422 & 6 & 16 & 1.091( 2) &      8.956/9  &      0.441 \\
 3 2 & 0.1467 & 11 & 16 & 0.894( 1) &      0.346/4  &      0.987 \\
 3 3 & 0.1525 & 10 & 16 & 0.644( 1) &      5.312/5  &      0.379 \\
 4 1 (WW)& 0.1442 & 6 & 16 & 1.015( 3) &     11.067/9  &      0.271 \\
 4 2 & 0.1487 & 11 & 16 & 0.816( 1) &      0.928/4  &      0.921 \\
 4 3 & 0.1545 & 10 & 16 & 0.552( 1) &      7.073/5  &      0.215 \\
 4 4 & 0.1565 & 5 & 16 & 0.447( 1) &     14.519/10  &      0.151 \\
 5 1 (WW)& 0.1452 & 6 & 16 & 0.978( 3) &     12.355/9  &      0.194 \\
 5 2 & 0.1497 & 10 & 16 & 0.776( 1) &      2.748/5  &      0.739 \\
 5 3 & 0.1555 & 5 & 16 & 0.502( 1) &     14.849/10  &      0.138 \\
 5 4 & 0.1575 & 5 & 16 & 0.393( 1) &     13.308/10  &      0.207 \\
 5 5 & 0.1585 & 5 & 16 & 0.331( 1) &     11.980/10  &      0.286 \\
 6 1 (WW)& 0.1460 & 6 & 16 & 0.952( 4) &     14.464/9  &      0.107 \\
 6 2 & 0.1505 & 10 & 16 & 0.748( 2) &      2.100/5  &      0.835 \\
 6 3 & 0.1562 & 5 & 16 & 0.467( 1) &     11.694/10  &      0.306 \\
 6 4 & 0.1583 & 5 & 16 & 0.350( 1) &      9.674/10  &      0.470 \\
 6 5 & 0.1593 & 5 & 16 & 0.280( 2) &      7.110/10  &      0.715 \\
 6 6 & 0.1600 & 4 & 16 & 0.214( 2) &      6.099/11  &      0.867 \\
\hline
\end{tabular}
\end{center}
\caption{
  Fits to pseudoscalar mesons, with Wilson valence fermions and
  $am_q=0.01$  staggered sea quarks. All fits are to ``kind=1'' WP propagators,
  unless otherwise indicated, and
  are to a single exponential.
  In this and following tables, numbers in the ``kind'' column
  for mesons refers to their quark content: 1 through 6 refer to
  hopping parameter .1320, .1410, .1525, .1565, .1585, and .1600.
}
\label{tab1}
\end{table}

\begin{table}
\begin{center}
\begin{tabular}{|l|c|r|r|l|r|r|} \hline
\multicolumn{1}{|c|}{kind} & $\kappa_{ave}$ & $D_{min}$ & $D_{max}$
  & \multicolumn{1}{|c|}{mass} & $\chi^2$/dof & C.L.\\
\hline
 1 1 (WW)& 0.1320 & 5 & 16 & 1.493( 2) &      9.056/10  &      0.527 \\
 2 1 (WW)& 0.1365 & 5 & 16 & 1.321( 2) &      7.275/10  &      0.699 \\
 2 2 & 0.1410 & 11 & 16 & 1.137( 1) &      0.591/4  &      0.964 \\
 3 1 (WW)& 0.1422 & 6 & 16 & 1.106( 3) &      5.972/9  &      0.743 \\
 3 2 & 0.1467 & 10 & 16 & 0.919( 1) &      1.689/5  &      0.890 \\
 3 3 & 0.1525 & 6 & 16 & 0.688( 1) &      6.147/9  &      0.725 \\
 4 1 (WW)& 0.1442 & 6 & 16 & 1.030( 3) &      5.885/9  &      0.751 \\
 4 2 & 0.1487 & 10 & 16 & 0.844( 2) &      1.565/5  &      0.905 \\
 4 3 & 0.1545 & 6 & 16 & 0.610( 2) &      4.314/9  &      0.890 \\
 4 4 & 0.1565 & 8 & 16 & 0.526( 3) &      5.088/7  &      0.649 \\
 5 1 (WW)& 0.1452 & 6 & 16 & 0.993( 4) &      5.877/9  &      0.752 \\
 5 2 & 0.1497 & 6 & 16 & 0.804( 1) &      7.780/9  &      0.556 \\
 5 3 & 0.1555 & 5 & 16 & 0.572( 2) &      8.688/10  &      0.562 \\
 5 4 & 0.1575 & 8 & 16 & 0.485( 3) &      6.766/7  &      0.454 \\
 5 5 & 0.1585 & 8 & 16 & 0.442( 4) &      8.550/7  &      0.287 \\
 6 1 (WW)& 0.1460 & 7 & 16 & 0.960( 6) &      4.150/8  &      0.843 \\
 6 2 & 0.1505 & 6 & 16 & 0.778( 2) &      6.007/9  &      0.739 \\
 6 3 & 0.1562 & 5 & 16 & 0.546( 3) &     13.120/10  &      0.217 \\
 6 4 & 0.1583 & 8 & 16 & 0.455( 4) &     12.455/7  &      0.087 \\
 6 5 & 0.1593 & 8 & 16 & 0.411( 6) &     13.128/7  &      0.069 \\
 6 6 & 0.1600 & 11 & 16 & 0.391(19) &      4.901/4  &      0.298 \\
\hline
\end{tabular}
\end{center}
\caption{
  Fits to vector mesons, with Wilson valence fermions and
  $am_q=0.01$  staggered sea quarks.
  All fits are to ``kind=1'' WP propagators, unless otherwise indicated,
  and are to a single exponential.
}
\label{tab2}
\end{table}

\begin{table}
\begin{center}
\begin{tabular}{|l|c|r|r|l|r|r|} \hline
\multicolumn{1}{|c|}{kind} & $\kappa_{ave}$ & $D_{min}$ & $D_{max}$
  & \multicolumn{1}{|c|}{mass} & $\chi^2$/dof & C.L.\\
\hline
(WW) & 0.1320 & 7 & 16 & 2.327(14) &      5.911/8  &      0.657 \\
 & 0.1410 & 11 & 16 & 1.777( 3) &      4.052/4  &      0.399 \\
 & 0.1525 & 9 & 16 & 1.097( 4) &      7.926/6  &      0.244 \\
 & 0.1565 & 8 & 16 & 0.839( 5) &      4.193/7  &      0.757 \\
 & 0.1585 & 8 & 16 & 0.699( 8) &      3.864/7  &      0.795 \\
 &  0.1600 & 7 & 16 & 0.610(23) &      8.984/8  &      0.344 \\
\hline
\end{tabular}
\end{center}
\caption{
  Fits to nucleons, with Wilson valence fermions and
  $am_q=0.01$  staggered sea quarks. All fits are to  WP propagators,
  unless otherwise indicated, and
  are to a single exponential.
}
\label{tab3}
\end{table}

\begin{table}
\begin{center}
\begin{tabular}{|l|c|r|r|l|r|r|} \hline
\multicolumn{1}{|c|}{kind} & $\kappa_{ave}$ & $D_{min}$ & $D_{max}$
  & \multicolumn{1}{|c|}{mass} & $\chi^2$/dof & C.L.\\
\hline
 (WW) & 0.1320 & 7 & 16 & 2.331(14) &      6.130/8  &      0.633 \\
  & 0.1410 & 11 & 16 & 1.786( 3) &      4.027/4  &      0.402 \\
  & 0.1525 & 10 & 16 & 1.123( 5) &      3.768/5  &      0.583 \\
  & 0.1565 & 8 & 16 & 0.876( 6) &      5.423/7  &      0.608 \\
  & 0.1585 & 5 & 16 & 0.743( 6) &     11.553/10  &      0.316 \\
  & 0.1600 & 4 & 16 & 0.628(10) &      4.594/11  &      0.949 \\
\hline
\end{tabular}
\end{center}
\caption{
  Fits to deltas, with Wilson valence fermions and
  $am_q=0.01$  staggered sea quarks.
  All fits are to ``kind=1'' WP propagators,
  unless otherwise indicated, and are to a single exponential.
}
\label{tab4}
\end{table}

\begin{table}
\begin{center}
\begin{tabular}{|l|c|r|r|l|r|r|} \hline
\multicolumn{1}{|c|}{kind} & $\kappa_{ave}$ & $D_{min}$ & $D_{max}$
  & \multicolumn{1}{|c|}{mass} & $\chi^2$/dof & C.L.\\
\hline
 1 1 (WW)& 0.1320 & 5 & 16 & 1.500( 2) &     13.391/10  &      0.203 \\
 2 1 (WW)& 0.1365 & 5 & 16 & 1.328( 2) &     12.433/10  &      0.257 \\
 2 2 & 0.1410 & 11 & 16 & 1.135( 1) &      3.975/4  &      0.409 \\
 3 1 (WW)& 0.1422 & 4 & 16 & 1.112( 2) &     22.669/11  &      0.020 \\
 3 2 & 0.1467 & 10 & 16 & 0.912( 1) &      2.320/5  &      0.803 \\
 3 3 & 0.1525 & 8 & 16 & 0.664( 1) &      1.299/7  &      0.988 \\
 4 1 (WW)& 0.1442 & 9 & 16 & 1.047( 4) &     16.354/6  &      0.012 \\
 4 2 & 0.1487 & 9 & 16 & 0.833( 1) &      3.820/6  &      0.701 \\
 4 3 & 0.1545 & 6 & 16 & 0.571( 1) &      4.799/9  &      0.851 \\
 4 4 & 0.1565 & 5 & 16 & 0.470( 1) &      5.376/10  &      0.865 \\
 5 1 (WW)& 0.1452 & 4 & 16 & 0.999( 2) &     27.140/11  &      0.004 \\
 5 2 & 0.1497 & 8 & 16 & 0.794( 1) &      4.970/7  &      0.664 \\
 5 3 & 0.1555 & 6 & 16 & 0.524( 1) &      3.550/9  &      0.938 \\
 5 4 & 0.1575 & 4 & 16 & 0.417( 1) &      6.493/11  &      0.839 \\
 5 5 & 0.1585 & 4 & 16 & 0.358( 1) &      8.052/11  &      0.709 \\
 6 1 (WW)& 0.1460 & 4 & 16 & 0.968( 2) &     20.236/11  &      0.042 \\
 6 2 & 0.1505 & 7 & 16 & 0.764( 1) &      6.954/8  &      0.542 \\
 6 3 & 0.1562 & 5 & 16 & 0.488( 1) &      5.677/10  &      0.842 \\
 6 4 & 0.1583 & 4 & 16 & 0.375( 1) &      7.843/11  &      0.727 \\
 6 5 & 0.1593 & 4 & 16 & 0.310( 1) &     12.291/11  &      0.342 \\
 6 6 & 0.1600 & 6 & 16 & 0.249( 2) &      8.813/9  &      0.455 \\
\hline
\end{tabular}
\end{center}
\caption{
  Fits to pseudoscalar mesons, with Wilson valence fermions and
  $am_q=0.025$  staggered sea quarks.
  All fits are to ``kind=1'' WP propagators,
  unless otherwise indicated, and are to a single exponential.
}
\label{tab5}
\end{table}

\begin{table}
\begin{center}
\begin{tabular}{|l|c|r|r|l|r|r|} \hline
\multicolumn{1}{|c|}{kind} & $\kappa_{ave}$ & $D_{min}$ & $D_{max}$
  & \multicolumn{1}{|c|}{mass} & $\chi^2$/dof & C.L.\\
\hline
 1 1 (WW)& 0.1320 & 5 & 16 & 1.510( 2) &     12.381/10  &      0.260 \\
 2 1 (WW)& 0.1365 & 5 & 16 & 1.340( 2) &     12.593/10  &      0.247 \\
 2 2 & 0.1410 & 10 & 16 & 1.153( 1) &      4.781/5  &      0.443 \\
 3 1 (WW)& 0.1422 & 4 & 16 & 1.129( 2) &     16.722/11  &      0.116 \\
 3 2 & 0.1467 & 9 & 16 & 0.939( 1) &      3.797/6  &      0.704 \\
 3 3 & 0.1525 & 7 & 16 & 0.716( 1) &      4.054/8  &      0.852 \\
 4 1 (WW)& 0.1442 & 4 & 16 & 1.056( 2) &     17.731/11  &      0.088 \\
 4 2 & 0.1487 & 9 & 16 & 0.867( 1) &      3.451/6  &      0.750 \\
 4 3 & 0.1545 & 6 & 16 & 0.638( 2) &      6.693/9  &      0.669 \\
 4 4 & 0.1565 & 6 & 16 & 0.560( 2) &      8.757/9  &      0.460 \\
 5 1 (WW)& 0.1452 & 4 & 16 & 1.019( 2) &     14.718/11  &      0.196 \\
 5 2 & 0.1497 & 7 & 16 & 0.830( 1) &      5.110/8  &      0.746 \\
 5 3 & 0.1555 & 6 & 16 & 0.601( 2) &      8.517/9  &      0.483 \\
 5 4 & 0.1575 & 8 & 16 & 0.520( 3) &      7.019/7  &      0.427 \\
 5 5 & 0.1585 & 4 & 16 & 0.483( 3) &     15.228/11  &      0.172 \\
 6 1 (WW)& 0.1460 & 4 & 16 & 0.991( 3) &      9.075/11  &      0.615 \\
 6 2 & 0.1505 & 7 & 16 & 0.805( 2) &      3.556/8  &      0.895 \\
 6 3 & 0.1562 & 8 & 16 & 0.574( 3) &      3.806/7  &      0.802 \\
 6 4 & 0.1583 & 8 & 16 & 0.494( 4) &      5.101/7  &      0.648 \\
 6 5 & 0.1593 & 4 & 16 & 0.458( 3) &     14.068/11  &      0.229 \\
 6 6 & 0.1600 & 4 & 16 & 0.434( 5) &     12.901/11  &      0.300 \\
\hline
\end{tabular}
\end{center}
\caption{
  Fits to vector mesons, with Wilson valence fermions and
  $am_q=0.025$  staggered sea quarks.
  All fits are to ``kind=1'' WP propagators,
  unless otherwise indicated, and are to a single exponential.
}
\label{tab6}
\end{table}

\begin{table}
\begin{center}
\begin{tabular}{|l|c|r|r|l|r|r|} \hline
\multicolumn{1}{|c|}{kind} & $\kappa$ & $D_{min}$ & $D_{max}$
  & \multicolumn{1}{|c|}{mass} & $\chi^2$/dof & C.L.\\
\hline
 (WW) & 0.1320 & 5 & 16 & 2.364(10) &      9.873/10  &      0.452 \\
 (WW) & 0.1410 & 5 & 16 & 1.833(10) &      7.799/10  &      0.649 \\
  & 0.1525 & 10 & 16 & 1.139( 5) &      6.934/5  &      0.226 \\
  & 0.1565 & 6 & 16 & 0.878( 4) &      9.811/9  &      0.366 \\
  & 0.1585 & 6 & 16 & 0.744( 6) &      8.209/9  &      0.513 \\
  & 0.1600 & 6 & 16 & 0.642(12) &      5.432/9  &      0.795 \\
\hline
\end{tabular}
\end{center}
\caption{
  Fits to nucleons, with Wilson valence fermions and
  $am_q=0.025$  staggered sea quarks. All fits are to  WP propagators,
  unless otherwise indicated, and are to a single exponential.
}
\label{tab7}
\end{table}

\begin{table}
\begin{center}
\begin{tabular}{|l|c|r|r|l|r|r|} \hline
\multicolumn{1}{|c|}{kind} & $\kappa_{ave}$ & $D_{min}$ & $D_{max}$
  & \multicolumn{1}{|c|}{mass} & $\chi^2$/dof & C.L.\\
\hline
 (WW) & 0.1320 & 5 & 16 & 2.368(10) &      8.734/10  &      0.558 \\
 (WW) & 0.1410 & 5 & 16 & 1.844(11) &      6.451/10  &      0.776 \\
  & 0.1525 & 10 & 16 & 1.164( 6) &      3.366/5  &      0.644 \\
  & 0.1565 & 6 & 16 & 0.918( 6) &      9.785/9  &      0.368 \\
  & 0.1585 & 6 & 16 & 0.804(10) &      8.522/9  &      0.483 \\
  & 0.1600 & 4 & 16 & 0.711( 9) &      8.337/11  &      0.683 \\
\hline
\end{tabular}
\end{center}
\caption{
  Fits to deltas, with Wilson valence fermions and
  $am_q=0.025$  staggered sea quarks.
  All fits are to ``kind=1'' WP propagators,
  unless otherwise indicated, and
  are to a single exponential.
}
\label{tab8}
\end{table}

\begin{table}
\begin{center}
\begin{tabular}{|l|c|r|r|l|r|r|} \hline
\multicolumn{1}{|c|}{kind} & $\kappa_{ave}$ & $D_{min}$ & $D_{max}$
  & mass ratio & $\chi^2$/dof & C.L. \\
\hline
  (WW)& 0.1320 & 12 & 16 & 0.997( 3) &      8.990/6  &      0.174 \\
  & 0.1410 & 11 & 16 & 0.985( 1) &      2.994/8  &      0.935 \\
  & 0.1525 & 11 & 16 & 0.933( 2) &      6.317/8  &      0.612 \\
  & 0.1565 & 11 & 16 & 0.853( 5) &     10.740/8  &      0.217 \\
  & 0.1585 & 11 & 16 & 0.752(11) &     11.370/8  &      0.182 \\
  & 0.1600 & 11 & 16 & 0.558(27) &      5.007/8  &      0.757 \\
\hline
\end{tabular}
\end{center}
\caption{
  Fits to the ratio $m_\pi/m_\rho$, with Wilson valence fermions and
  $am_q=0.01$  staggered sea quarks.
  All fits are to ``kind=1'' WP propagators,
  unless otherwise indicated, and
  are correlated fits to a single exponential in each channel.
}
\label{tab9}
\end{table}

\begin{table}
\begin{center}
\begin{tabular}{|l|c|r|r|l|r|r|} \hline
\multicolumn{1}{|c|}{kind} & $\kappa$ & $D_{min}$ & $D_{max}$
  & mass ratio & $\chi^2$/dof & C.L.\\
\hline
   (WW)& 0.1320 & 7 & 16 & 1.563( 8) &     11.600/16  &      0.771 \\
  & 0.1410 & 11 & 16 & 1.563( 3) &      4.961/8  &      0.762 \\
  & 0.1525 & 8 & 16 & 1.590( 5) &     12.880/14  &      0.536 \\
  & 0.1565 & 7 & 16 & 1.591(11) &     12.520/16  &      0.707 \\
  & 0.1585 & 8 & 16 & 1.597(30) &     17.370/14  &      0.237 \\
  & 0.1600 & 6 & 16 & 1.573(53) &     39.600/18  &      0.002 \\
\hline
\end{tabular}
\end{center}
\caption{
  Fits to the ratio $m_N/m_\rho$, with Wilson valence fermions and
  $am_q=0.01$  staggered sea quarks.
  All fits are to ``kind=1'' WP propagators,
  unless otherwise indicated, and
  are correlated fits to a single exponential in each channel.
}
\label{tab10}
\end{table}

\begin{table}
\begin{center}
\begin{tabular}{|l|c|r|r|l|r|r|} \hline
\multicolumn{1}{|c|}{kind} & $\kappa_{ave}$ & $D_{min}$ & $D_{max}$
  & mass ratio & $\chi^2$/dof & C.L.\\
\hline
 (WW) & 0.1320 & 13 & 16 & 1.002( 5) &     14.510/4  &      0.006 \\
 (WW) & 0.1410 & 6 & 16 & 0.984( 2) &     25.800/18  &      0.104 \\
  & 0.1525 & 11 & 16 & 0.984( 1) &      3.694/8  &      0.884 \\
  & 0.1565 & 7 & 16 & 0.929( 1) &      7.176/16  &      0.970 \\
  & 0.1585 & 7 & 16 & 0.845( 3) &     14.150/16  &      0.588 \\
  & 0.1600 & 8 & 16 & 0.744( 7) &     14.560/14  &      0.409 \\
\hline
\end{tabular}
\end{center}
\caption{
  Fits to $m_\pi/m_\rho$, with Wilson valence fermions and
  $am_q=0.025$  staggered sea quarks.
  All fits are to ``kind=1'' WP propagators,
  unless otherwise indicated, and
  are  correlated fits to a single exponential in each channel.
}
\label{tab11}
\end{table}

\begin{table}
\begin{center}
\begin{tabular}{|l|c|r|r|l|r|r|} \hline
\multicolumn{1}{|c|}{kind} & $\kappa_{ave}$ & $D_{min}$ & $D_{max}$
  & mass ratio & $\chi^2$/dof & C.L.\\
\hline
 (WW) & 0.1320 & 6 & 16 & 1.564( 8) &     23.060/18  &      0.188 \\
 (WW) & 0.1410 & 5 & 16 & 1.584( 9) &     30.380/20  &      0.064 \\
  & 0.1525 & 11 & 16 & 1.568( 2) &      8.207/8  &      0.414 \\
  & 0.1565 & 11 & 16 & 1.596( 5) &      5.214/8  &      0.734 \\
  & 0.1585 & 6 & 16 & 1.553( 9) &     21.590/18  &      0.251 \\
  & 0.1600 & 6 & 16 & 1.516(15) &     21.200/18  &      0.269 \\
\hline
\end{tabular}
\end{center}
\caption{
  Fits to $m_N/m_\rho$, with Wilson valence fermions and
  $am_q=0.025$  staggered sea quarks.
  All fits are to ``kind=1'' WP propagators,
  unless otherwise indicated, and
  are correlated fits to a single exponential in each channel.
}
\label{tab12}
\end{table}

\begin{table}
\begin{center}
\begin{tabular}{|l|c|l|} \hline
 $am_q$  & particle  & \multicolumn{1}{|c|}{mass}   \\
\hline
0.01 & $\rho$ & .360(8) \\
     & p      & .524(18) \\
     & $\Delta$ & .560(12) \\
0.025 & $\rho$   & .386(5) \\
      & p        & .558(12) \\
      & $\Delta$ & .651(17) \\
\hline
\end{tabular}
\end{center}
\caption{
  Extrapolations to $\kappa_c$.
}
\label{tab13}
\end{table}

\begin{table}
\begin{center}
\begin{tabular}{|l|c|r|r|l|r|r|} \hline
\multicolumn{1}{|c|}{kind} & $\kappa_{ave}$ & $D_{min}$ & $D_{max}$
  & \multicolumn{1}{|c|}{mass} & $\chi^2$/dof & C.L.\\
\hline
 pion & 0.1585 & 7 & 16 & 0.378( 2) &     12.382/8  &      0.135 \\
  & 0.1600 & 7 & 16 & 0.273( 3) &     12.821/8  &      0.118 \\
 rho & 0.1585 & 8 & 16 & 0.530( 6) &      2.857/7  &      0.898 \\
  & 0.1600 & 7 & 16 & 0.486( 9) &      5.388/8  &      0.715 \\
 proton & 0.1585 & 7 & 16 & 0.783(10) &      8.339/8  &      0.401 \\
  & 0.1600 & 6 & 16 & 0.673( 9) &      8.113/9  &      0.523 \\
 delta & 0.1585 & 8 & 16 & 0.852(11) &      9.302/7  &      0.232 \\
  & 0.1600 & 8 & 16 & 0.757(25) &      3.628/7  &      0.821 \\
\hline
\end{tabular}
\end{center}
\caption{
  Fits to quenched $\beta=5.85$ spectroscopy with
  Wilson valence fermions. All fits
  are to a single exponential.
}
\label{tab14}
\end{table}

\begin{table}
\begin{center}
\begin{tabular}{|l|c|r|r|l|r|r|} \hline
\multicolumn{1}{|c|}{kind} & $\kappa_{ave}$ & $D_{min}$ & $D_{max}$
  & \multicolumn{1}{|c|}{mass} & $\chi^2$/dof & C.L.\\
\hline
 pion & 0.1554 & 7 & 16 & 0.362( 1) &      5.284/8  &      0.727 \\
  & 0.1567 & 7 & 16 & 0.271( 2) &     10.564/8  &      0.228 \\
 rho & 0.1554 & 5 & 16 & 0.486( 3) &     13.433/10  &      0.200 \\
  & 0.1567 & 4 & 16 & 0.445( 5) &     19.847/11  &      0.047 \\
 proton & 0.1554 & 6 & 16 & 0.721( 6) &     11.226/9  &      0.261 \\
  & 0.1567 & 6 & 16 & 0.619( 8) &      4.814/9  &      0.850 \\
 delta & 0.1554 & 6 & 16 & 0.777( 7) &     12.580/9  &      0.183 \\
  & 0.1567 & 6 & 16 & 0.699( 9) &      7.574/9  &      0.578 \\
\hline
\end{tabular}
\end{center}
\caption{
  Fits to quenched $\beta=5.95$ spectroscopy with
  Wilson valence fermions. All fits
  are to a single exponential.
}
\label{tab15}
\end{table}

\begin{table}
\begin{center}
\begin{tabular}{|l|c|r|r|l|r|r|} \hline
\multicolumn{1}{|c|}{kind} & $\kappa$ & $D_{min}$ & $D_{max}$
  & mass ratio & $\chi^2$/dof & C.L.\\
\hline
 $m_\pi/m_\rho$ & 0.1585 & 7 & 16 & 0.716( 6) & 11.050/16  &   0.806 \\
              & 0.1600 & 7 & 16 & 0.573(10) &  11.740/16  &      0.762 \\
 $m_N/m_\rho$ & 0.1585 & 7 & 16 & 1.468(16) &    8.603/16  &      0.929\\
              & 0.1600 & 5 & 16 & 1.415(24) &   15.060/20  &      0.773 \\
\hline
\end{tabular}
\end{center}
\caption{
  Fits to ratios $m_\pi/m_\rho$ and $m_N/m_\rho$,
  from quenched $\beta=5.85$ simulations
  with Wilson valence fermions. All fits are to ``kind=1'' WP propagators,
  and are correlated fits to a single exponential in each channel.
}
\label{tab16}
\end{table}

\begin{table}
\begin{center}
\begin{tabular}{|l|c|r|r|l|r|r|} \hline
\multicolumn{1}{|c|}{kind} & $\kappa_{ave}$ & $D_{min}$ & $D_{max}$
  & mass ratio & $\chi^2$/dof & C.L.\\
\hline
 $m_\pi/m_\rho$ & 0.1554 & 10 & 16 & 0.740( 8) &  5.786/10  &  0.833 \\
                & 0.1567 & 10 & 16 & 0.599(16) &  7.037/10  &  0.722 \\
 $m_N/m_\rho$   & 0.1554 & 10 & 16 & 1.507(27) &  7.394/10  &    0.688 \\
                & 0.1567 & 4 & 16 & 1.420(18) &  24.370/22  &   0.328\\
\hline
\end{tabular}
\end{center}
\caption{
  Fits to ratios $m_\pi/m_\rho$ and $m_N/m_\rho$,
  from quenched $\beta=5.95$ simulations
  with Wilson valence fermions.
  All fits are to ``kind=1'' WP propagators,
  and are correlated fits to a single exponential in each channel.
}
\label{tab17}
\end{table}

\end{document}